\documentclass[aps,twocolumn,showpacs,preprintnumbers,floats]{revtex4}\def\@cite#1#2{\textsuperscript{[{#1\if@tempswa , #2\fi}]}}
\usepackage{mathrsfs}
\usepackage{longtable,lscape}
\usepackage{txfonts}
\usepackage{amssymb}
\usepackage{indentfirst}
\usepackage{graphicx,,booktabs}
\usepackage{color}
\usepackage{amssymb}
\usepackage{epsfig}

\begin{document}
\title{Low energy reaction $K^-p\rightarrow\Lambda\eta$ and the negative parity $\Lambda$ resonances }
\author{
Li-Ye Xiao and Xian-Hui Zhong \footnote {E-mail:
zhongxh@hunnu.edu.cn}} \affiliation{ Department of Physics, Hunan
Normal University, and Key Laboratory of Low-Dimensional Quantum
Structures and Quantum Control of Ministry of Education, Changsha
410081, China }

%\date{\today}

\begin{abstract}

The reaction $K^-p\rightarrow\Lambda\eta$ at low energies is studied
with a chiral quark model approach. Good descriptions of the
existing experimental data are obtained. It is found that
$\Lambda(1670)$ dominates the reaction around threshold.
Furthermore,  $u$- and $t$-channel backgrounds play crucial roles in
this reaction as well. The contributions from the $D$-wave state
$\Lambda(1690)$ are negligibly small for its tiny coupling to
$\eta\Lambda$. To understand the strong coupling properties of the
low-lying negative parity $\Lambda$ resonances extracted from the
$\bar{K}N$ scattering, we further study their strong decays. It is
found that these resonances are most likely mixed states between
different configurations. Considering these low-lying negative
parity $\Lambda$ resonances as mixed three-quark states, we can
reasonably understand both their strong decay properties from
Particle Data Group and their strong coupling properties extracted
from the $\bar{K}N$ scattering. As a byproduct, we also predict the
strong decay properties of the missing $D$-wave state
$|\Lambda\frac{3}{2}^-\rangle_3$ with a mass of $\sim1.8$ GeV. We
suggest our experimental colleagues search it in the
$\Sigma(1385)\pi$ and $\Sigma\pi$ channels.

\end{abstract}


\maketitle

\section{INTRODUCTION}

Our knowledge about $\Lambda$ resonances is much poorer than that of
nucleon resonances~\cite{Beringer:2012}. Even for the
well-established low-lying negative parity states, such as
$\Lambda(1405)S_{01}$, $\Lambda(1520)D_{03}$ and
$\Lambda(1670)S_{01}$, their properties are still
controversial~\cite{Klempt:2009pi}. Up to now we can not clarify
whether these states are excited three quark states, dynamically
generated resonances, three quark states containing multi-quark
components or the other explanations, although there are extensive
discussions about their
natures~\cite{Chen:2009de,Isgur78,Isgur:1977ky,Capstick:1986bm,Melde:2008,
Bijker:2000, Glozman:1997ag,Loring:2001ky,Schat:2001xr,
Goity:2002pu,Menadue:2011pd,Engel:2012qp,Koniuk:1979vy,Melde:2006yw,Melde:2007,An:2010wb,
Hyodo:2011ur,Oller:2005ig,Roca:2006sz,
Oller:2006hx,Borasoy:2005ie,Hyodo:2003qa,Oset:2001cn,
GarciaRecio:2002td,Jido:2003cb,Oller:2000fj,
Oset:1997it,Oller:2006jw,Borasoy:2006sr,Roca:2008kr,Bouzasa;2008,Martin:1969ud,Manley:2002,
Lutz:2001yb, Buttgen:1985yz,MuellerGroeling:1990cw,Hamaie:1995wy,
Zhong:2009,Martin:1980qe,Gensini:1997fp,Guo:2012vv,Zhang:2013cua,Zhang:2013sva,
Geng:2007vm,Geng:2007hz,Geng:2008er,Liu:2012ge,Liu:2011sw,Xie:2013wfa,GarciaRecio:2003ks,Zou:2008be}.

Recently, we systematically studied the reactions $K^-p\rightarrow
\Sigma^0\pi^0,\Lambda\pi^0,\bar{K}^0n$ in a chiral quark model
approach~\cite{Zhong:2013oqa}. Obvious roles of the low-lying
negative parity states, $\Lambda(1405)$, $\Lambda(1520)$ and
$\Lambda(1670)$, are found in the $K^-p\rightarrow
\Sigma^0\pi^0,\bar{K}^0n$ reactions, where we have extracted their
strong coupling properties. For example, we found that
$\Lambda(1670)$ should have a very weak coupling to $\bar{K}N$,
while $\Lambda(1520)$ needs a strong coupling to $\bar{K}N$, which
can not be well explained with the symmetry constituent quark model
in the $SU(6)\otimes O(3)$ limit~\cite{Zhong:2013oqa}.

To obtain more strong coupling properties and better understandings
of these low-lying $\Lambda$ resonances, in this work, we will
continue to study another important $\bar{K}N$ reaction
$K^-p\rightarrow \eta\Lambda$. This reaction provides us a very
clear place to study the low-lying $\Lambda$ resonances, because
only the $\Lambda$ resonances contribute here due to the isospin
selection rule. Especially, the poorly known strong coupling of
$\Lambda(1670)$ to $\eta\Lambda$ might be reliably obtained from the
$K^-p\rightarrow \eta\Lambda$, for this reaction at threshold is
dominated by formation of the
$\Lambda(1670)$~\cite{Starostin:2001zz}. Recently, the new data of
the $K^-p\rightarrow \eta\Lambda$ reaction from Crystal Ball
Collaboration~\cite{Starostin:2001zz} were analyzed with an
effective Lagrangian model by Liu and
Xie~\cite{Liu:2012ge,Liu:2011sw}. They might find some evidence of a
exotic $D$-wave resonance with mass $M\simeq 1669$ MeV and width
$\Gamma\simeq 1.5$ MeV in the reaction, which will be discussed in
present work as well.

Furthermore, to understand the natures of these strong coupling
properties extracted from the $\bar{K}N$ scattering, we will further
carry out a systematical study of the strong decays of the low-lying
negative parity $\Lambda$ resonances in the chiral quark model
approach as well. Combing the strong coupling properties extracted
from the $\bar{K}N$ scattering with the strong decay properties from
the Particle Data Group (PDG)~\cite{Beringer:2012}, we expect to
obtain more reliable understandings of the natures for these
low-lying negative parity $\Lambda$ resonances.

This work is organized as follows. In Sec.~\ref{FRAME}, the model is
reviewed. Then, the numerical results are presented and discussed in
Sec.~\ref{RESULT}. Finally, a summary is given in Sec.~\ref{summ}.

\section{FRAMEWORK}\label{FRAME}

In this work, we study the $K^-p\rightarrow \eta\Lambda$ reaction in
a chiral quark model. This model has been well developed and widely
applied to meson photoproduction
reactions~\cite{qkk,Li:1997gda,zhao-kstar,qk3,qk4,qk5,He:2008ty,Saghai:2001yd,Zhong:2011ht}.
Its recent extension to describe the $\pi N$ ~\cite{Zhong:2007fx}
and $\bar{K} N$ ~\cite{Zhong:2009,Zhong:2013oqa} reactions also
turns out to be successful and inspiring.

In the calculations, we consider three basic Feynman diagrams, i.e.,
$s$-, $u$- and $t$-channels at the tree level. The reaction
amplitude is expressed as
\begin{equation}
\mathcal{M}=\mathcal{M}_s+\mathcal{M}_u+\mathcal{M}_t,
\end{equation}
where the $s$- and $u$-channel reaction amplitudes $\mathcal{M}_s$
and $\mathcal{M}_u$ are given by
 \begin{eqnarray}
 \mathcal{M}_s=\sum_j\langle N_f|H^f_m|N_j\rangle\langle
 N_j|\frac{1}{E_i+\omega_i-E_j}H^i_m|N_i\rangle,\label{sc}\\
 \mathcal{M}_u=\sum_j\langle N_f|H^i_m\frac{1}{E_i-\omega_f-E_j}|N_j\rangle\langle
 N_jH^f_m|N_i\rangle.\label{uc}
 \end{eqnarray}
In the above equations, $H_m$ stands for the quark-meson coupling,
which might be described by the effective chiral
Lagrangian~\cite{qkk,Li:1997gda}
\begin{equation}
H_m=\sum_j\frac{1}{f_m}\overline{\psi}_j\gamma^j_u\gamma^j_5
\psi_j\vec{\tau}\cdot\partial^u\vec{\phi}_m,
\end{equation}
where $\psi_j$ represents the $j$th quark field in a baryon, and
$f_m$ is the meson's decay constant. The pseudoscalar meson octet
$\phi_m$ is written as
\begin{equation}
\phi_m=\left(\begin{array}{ccc}
\frac{1}{\sqrt{2}}\pi^0+\frac{1}{\sqrt{6}}\eta & \pi^+ & K^+ \cr
 \pi^- & -\frac{1}{\sqrt{2}}\pi^0+\frac{1}{\sqrt{6}}\eta & K^0 \cr
 K^- & \bar{K}^0 & -\sqrt{\frac{2}{3}}\eta
\end{array}\right).
\end{equation}
In Eqs. (\ref{sc}) and (\ref{uc}), $\omega_i$ and $\omega_f$ are the
energies of the incoming and outgoing mesons, respectively.
$|N_i\rangle$, $|N_j\rangle$ and $|N_f\rangle$ stand for the
initial, intermediate and final states, respectively, and their
corresponding energies are $E_i$, $E_j$ and $E_f$, which are the
eigenvalues of the nonrelativistic Hamiltonian of constituent quark
model $\hat{H}$~\cite{Isgur78, Isgur:1977ky,Capstick:1986bm}.

The resonance transition amplitudes of the $s$-channel can be
generally expressed as~\cite{Zhong:2007fx}
\begin{eqnarray}
\mathcal{M}^s_R=\frac{2M_R}{s-M^2_R+iM_R
\Gamma_R}\mathcal{O}_Re^{-(\textbf{k}^2+\textbf{q}^2)/(6\alpha^2)},\label{stt}
\end{eqnarray}
where $M_R$ and $\Gamma_R$ stand for the mass and width of the
resonance, respectively. The Mandelstam variable $s$ is defined as
$s\equiv(P_i+k)^2 $.  The single-resonance-excitation amplitude,
$\mathcal{O}_R$, can be obtain by the relation~\cite{Zhong:2013oqa}
\begin{eqnarray}\label{pt}
\mathcal{O}(n,l,J)=\sum_R\mathcal{O}_R(n,l,J)=\sum_Rg_R\mathcal{O}(n,l,J),
\end{eqnarray}
where $g_R$ stands for the relative strength of a single-resonance
in the partial amplitude $\mathcal{O}(n,l,J)$. The $g_R$ factors are
determined by the structure of each resonance and their couplings to
the meson and baryon. The partial amplitudes, $\mathcal{O}(n,l,J)$,
up to $n=2$ shell have been derived in our previous
work~\cite{Zhong:2013oqa}, where the details can be found. For
example, the important partial amplitude for the $S$ waves is given
by~\cite{Zhong:2013oqa}
\begin{eqnarray}
\mathcal{O}_1(S)&=&\left(g_{s1}-\frac{1}{2}g_{s2}\right)\Big(|\mathbf{A}_{out}|\cdot|\mathbf{A}_{in}
|\frac{|\mathbf{k}||\mathbf{q}|}{9\alpha^2}
+\frac{\omega_i}{6\mu_q}\mathbf{A}_{out}\cdot
\mathbf{q}\nonumber\\
&&+\frac{\omega_f}{6\mu_q}\mathbf{A}_{in}\cdot
\mathbf{k}+\frac{\omega_i\omega_f}{4\mu_q\mu_q}\alpha^2\Big),
\end{eqnarray}
where $\mathbf{k}$ and $\mathbf{q}$ stand for the three-momenta of
the incoming and outgoing mesons, respectively, and $\alpha$ is the
harmonic oscillator parameter. The reduced mass $\mu_q$ at the quark
level is defined by $1/\mu_q=1/m_q^i+1/m_q^f$, where $m_q^i$ and
$m_q^f$ correspond to the initial and final quark masses,
respectively. $\mathbf{A}_{in}$ and $\mathbf{A}_{out}$ are the same
variables defined in~\cite{Zhong:2013oqa}. The $g$-factors in the
partial amplitudes, such as $g_{s1}$ and $g_{s2}$, have been defined
in~\cite{Zhong:2013oqa} as well. These $g$-factors can be derived in
the SU(6)$\otimes$O(3) symmetry limit. In Tab.~\ref{abf}, we have
listed the $g$-factors for the reaction
$K^-p\rightarrow\eta\Lambda$.

And the transition amplitudes of the $u$-channel are given
by~\cite{Zhong:2007fx,Zhong:2009}
\begin{eqnarray}
\mathcal{M}^u_n=-\frac{2M_n}{u-M^2_n}\mathcal{O}_n
e^{-(\textbf{k}^2+\textbf{q}^2)/(6\alpha^2)}.\label{utt}
\end{eqnarray}
In Eq.(\ref{utt}), the amplitude $\mathcal{O}_n$ is determined by
the structure of each resonance and their couplings to the meson and
baryon, which has been derived in our previous
work~\cite{Zhong:2013oqa}. The Mandelstam variable $u$ are defined
as $u\equiv(P_i-q)^2 $.

In the calculations, we consider the vector- and scalar-exchanges
for the $t$-channel backgrounds. The vector meson-quark and scalar
meson-quark couplings are given by
\begin{eqnarray}\label{coup}
H_V&=& \bar{\psi}_j\left(a\gamma^{\nu}+\frac{
b\sigma^{\nu\lambda}\partial_{\lambda}}{2m_q}\right)V_{\nu} \psi_j,\\
H_S&=&g_{Sqq}\bar{\psi}_j\psi_jS,
\end{eqnarray}
where $V$ and $S$ stand for the vector and scalar fields,
respectively. The constants $a$, $b$ and $g_{Sqq}$ are the vector,
tensor and scalar coupling constants, respectively. They are treated
as free parameters in this work.

On the other hand, the $VPP$ and $SPP$ couplings ($P$ stands for a
pseudoscalar-meson) are adopted as
\begin{eqnarray}
H_{VPP}&=&-iG_VTr([\phi_m,\partial_\mu\phi_m]V^{\mu}),\\
H_{SPP}&=&\frac{g_{SPP}}{2m_\pi}\partial_\mu\phi_m\partial^\mu
\phi_m S,
\end{eqnarray}
where $G_V$ and $g_{SPP}$ are the $VPP$ and $SPP$ coupling constants
to be determined by experimental data.

For the vector meson exchange, the $t$-channel amplitude in the
quark model is written as~\cite{Zhong:2013oqa}
\begin{equation}
\mathcal{M}^V_t=\mathcal{O}^t_V\frac{1}{t-M^2_V}e^{-(\mathbf{q}-\mathbf{k})^2/(6\alpha^2)},
\end{equation}
where $e^{-(\mathbf{q}-\mathbf{k})^2/(6\alpha^2)}$ is a form factor
deduced from the quark model, and $M_V$ is the vector-meson mass.
The amplitude $\mathcal{O}^t_V$ is given by~\cite{Zhong:2013oqa}
\begin{eqnarray}
\mathcal{O}^t_V&=&G_va[g^s_t(\mathcal{H}_0+\mathcal{H}_1\mathbf{q}\cdot\mathbf{k})
+g^v_t\mathcal{H}_2i\mathbf{\sigma}\cdot(\mathbf{q}\times\mathbf{k})]\nonumber\\
&&+\text{tensor term},\label{tvector}
\end{eqnarray}
where the factors $g^s_t$ and $g^v_t$ are defined by $g^s_t\equiv
\langle N_f|\sum^3_{j=1}I^{ex}_j|N_i\rangle$ and $g^v_t\equiv
\langle N_f|\sum^3_{j=1}\sigma_j I^{ex}_j|N_i\rangle$, where,
$I^{ex}_j$ is the isospin operator of exchanged meson. These factors
can be deduced from the quark model.

For the scalar meson exchange, the $t$-channel amplitude in the
quark model is given by~\cite{Zhong:2013oqa}
\begin{equation}
\mathcal{M}^S_t=\mathcal{O}^t_S\frac{1}{t-M^2_S}e^{-(\mathbf{q}-\mathbf{k})^2/(6\alpha^2)},
\end{equation}
where $M_S$ is the scalar-meson mass, and the $\mathcal{O}^t_S$ is
written as~\cite{Zhong:2013oqa}
\begin{eqnarray}
\mathcal{O}^t_S&\simeq&\frac{g_{SPP}g_{Sqq}}{2m_\pi}(\omega_i\omega_f
-\mathbf{q}\cdot\mathbf{k})[g^s_t(\mathcal{A}_0+\mathcal{A}_1\mathbf{q}\cdot\mathbf{k})\nonumber\\
&&g^v_t\mathcal{A}_2i\mathbf{\sigma}\cdot(\mathbf{q}\times\mathbf{k})].\label{tscalar}
\end{eqnarray}
In Eqs.(\ref{tvector}) and (\ref{tscalar}), the variables
$\mathcal{H}_i$ and $\mathcal{A}_i$ ($i=0,1,2$) are the same
definitions as in ~\cite{Zhong:2013oqa}.

In this work, we consider the $K^{*}$- and $\kappa$-exchanges in the
$K^-p\rightarrow\Lambda\eta$ process. The factors $g^s_t$ and
$g^v_t$ derived from the quark model have been listed in
Tab.~\ref{abf} as well.

\begin{table}[ht]
\caption{The $g$-factors appearing in the $s$-, $u$- and $t$-channel
amplitudes of the $K^-p\rightarrow\Lambda\eta$ process obtained in
in the SU(6)$\otimes$O(3) symmetry limit. $\phi_p$ is the
$\eta$-$\eta'$ mixing angle defined
in~\cite{DiDonato:2011kr,Ke:2009mn}.}\label{abf}
\begin{tabular}{|c|c|c|c|c|c|c }\hline
 \hline
$g_{s1}=-\frac{\sqrt{6}}{6}\sin{\phi_p}$ &$g_{v1}=-\frac{\sqrt{6}}{4}\sin{\phi_p}$ \\
$g^u_{s1}=\frac{\sqrt{3}}{2}\cos{\phi_p}$  &$g^u_{v1}=\frac{\sqrt{3}}{2}\cos{\phi_p}$\\
$g^s_t=\frac{\sqrt{6}}{2}$ &$g^v_t=\frac{\sqrt{6}}{2}$\\
\hline
\end{tabular}
\end{table}

\section{RESULT AND ANALYSIS}\label{RESULT}

\subsection{Parameters}

With the transition amplitudes derived within the quark model, the
differential cross section can be calculated by~\cite{Zhong:2013oqa}
\begin{eqnarray}
\frac{d\sigma}{d\Omega}&=&\frac{(E_i+M_i)(E_f+M_f)}{64\pi^2s(2M_i)(2M_f)}
\frac{|\mathbf{q}|}{|\mathbf{k}|}\frac{M_N^2}{2}\nonumber\\
&&\times\sum_{\lambda_i,\lambda_f}\left|[\frac{\delta_{m_i}}{f_{m_i}}\frac{\delta_{m_f}}{f_{m_f}}
(\mathcal{M}_s+\mathcal{M}_u)+\mathcal{M}_t]_{\lambda_f,\lambda_i}\right|^2,
\end{eqnarray}
where $\lambda_i=\pm1/2$ and $\lambda_f=\pm1/2$ are the helicities
of the initial and final state $\Lambda$ baryons, respectively.
$f_{m_i}$ and $f_{m_f}$ are the initial and final meson decay
constants, respectively. $\delta_{m_i}\delta_{m_f}$ is a global
parameter accounting for the flavor symmetry breaking effects
arising from the quark-meson couplings, which will be determined by
experimental data.

In the calculation, the universal value of harmonic oscillator
parameter $\alpha=0.4$ GeV is adopted. The masses of the $u$, $d$,
and $s$ constituent quarks are set as $m_u=m_d=330$ MeV, and
$m_s=450$ MeV, respectively. The decay constants for $\eta$ and $K$
are adopted as $f_\eta=f_K=160$ MeV.

In present work, the resonance transition amplitude,
$\mathcal{O}_R$, is derived in the $SU(6)\otimes O(3)$ symmetric
quark model limit. In reality, due to e.g. spin-dependent forces in
the quark-quark interaction, the symmetry of $SU(6)\otimes O(3)$ is
generally broken. As a result, configuration mixing would
occur~\cite{Isgur78,Isgur:1977ky,Capstick:1986bm,Schat:2001xr}. To
take into account the breaking of that symmetry, a set of coupling
strength parameters, $C_R$, should be introduced for each resonance
amplitude,
\begin{equation}
\mathcal{O}_R\rightarrow C_R\mathcal{O}_R,
\end{equation}
where $C_R$ should be determined by fitting the experimental
observation. One finds that $C_R=1$ in the $SU(6)\otimes O(3)$
symmetry limit, while deviation of $C_R$ from unity implies the
$SU(6)\otimes O(3)$ symmetry breaking. The determined values of
$C_R$ for the $K^-p\rightarrow\Lambda\eta$ process have been listed
in Table~\ref{Pra}. These strength parameters $C_R$ for the main
resonances will be further discussed in Sec.~\ref{cxx}.

\begin{table}[ht]
\caption{The determined values for the parameters $C_R$,
$\delta_{m_i}\delta_{m_f}$ and $\phi_P$ in the
$K^-p\rightarrow\Lambda\eta$ scatting process.} \label{Pra}
\begin{tabular}{|c|c|c|c|c|c|c|c|c|c|c }\hline\hline
Parameter  & $C_{S_{01}(1405)}$ &$C_{D_{03}(1520)}$&$C_{S_{01}(1670)}$&$C_{D_{03}(1690)}$&$\delta_{m_i}\delta_{m_f}$&$\phi_P$\\
\hline
 Value  & 1.17 &  1.18 &34.70& 38.58 &1.24&$35^\circ$\\
\hline
\end{tabular}
\end{table}

In the $t$ channel, there are two parameters, $G_{V}a$ and
$g_{SPP}g_{Sqq}$, which come from $K^{*}$- and $\kappa$-exchanges,
respectively. By fitting the data, we obtain $G_Va\simeq4.8$ and
$g_{SPP}g_{Sqq}\simeq105$, which are consistent with our previous
determinations in~\cite{Zhong:2013oqa}.

In the $u$ channel, the intermediate states are nucleon and nucleon
resonances. One finds that the contributions from $n\geq1$ shell are
negligibly small, and are insensitive to the degenerate masses for
these shells. In present work, we take $M_1=1650$ MeV and $M_2=1750$
MeV for the degenerate masses of $n=1$ and $n=2$ shell nucleon
resonances, respectively.

In the $s$-channel of the $K^-p\rightarrow\Lambda\eta$ process,
there are only the contributions from $\Lambda$ resonances. The
low-lying $\Lambda$ resonances classified in the quark model up to
$n=2$ shell are listed in Tab.~\ref{qc}. From the table, we can see
that in $n=0$ shell, only the $\Lambda$ pole contribute to the
process, while in $n=1$ shell, two $S$-waves (i.e.,
$[70,^21]\Lambda(1405)S_{01}$, $[70,^28]\Lambda(1670)S_{01}$) and
two $D$-waves (i.e., $[70,^21]\Lambda(1520)D_{03}$,
$[70,^28]\Lambda(1690)D_{03}$) contribute to the reaction. The
excitations of $[70,^48]$ are forbidden for the $\Lambda$-selection
rule~\cite{Zhao:2006an}. In the calculations, the $n=2$ shell
$\Lambda$ resonances in the $s$ channel are treated as degeneration,
and their degenerate mass and width are taken as $M$=1800 MeV and
$\Gamma$=100 MeV, since in the low-energy region the contributions
from the $n =2$ shell are not significant.

By fitting the experimental data, we obtain their Breit-Wigner
masses and widths, which are listed in Tab.~\ref{BW}. From the
table, it is seen that the extracted resonances' parameters are
compatible with the data from PDG~\cite{Beringer:2012}.

\begin{table}[ht]
\caption{The classification of $\Lambda$ resonances in the quark
model up to n=2 shell. The "?" denotes the resonances being
unestablished. $l_{I,2J}$ is the PDG notation of baryons. $N_6$ and
$N_3$ denote the SU(6) and SU(3) representation. $L$ and $S$ stand
for the total orbital momentum and spin of a baryon, respectively.}
\label{qc}
\begin{tabular}{|c|c||c|c|c|c|c }\hline
\hline
$|N_6,^{2S+1}N_3,n,L\rangle$ &$l_{I,2J}$ &$|N_6,^{2S+1}N_3,n,L\rangle$ &$l_{I,2J}$\\
\hline
$|56,^28,0,0\rangle$ &$P_{01}(1116)$&...&...\\
\hline
$|70,^21,1,1\rangle$ &$S_{01}(1405)$  &$|56,^28,2,2\rangle$   &$P_{03}(?)$\\
                     &$D_{03}(1520)$  &                       &$F_{05}(?)$\\
\hline
$|70,^28,1,1\rangle$ &$S_{01}(1670)$  &$|70,^21,2,2\rangle$  &$P_{03}(?)$\\
                     &$D_{03}(1690)$  &                      &$F_{05}(?)$\\
\hline
$|70,^48,1,1\rangle$  &$S_{01}(1800)$  &$|70,^28,2,2\rangle$  &$P_{03}(?)$\\
                      &$D_{03}(?)$     &                      &$F_{05}(?)$\\
                      &$D_{05}(1830)$  &                       &    \\
\hline
$|56,^28,2,0\rangle$ &$P_{01}(1600)$  &$|70,^48,2,2\rangle$ &$P_{01}(?)$\\
$|70,^21,2,0\rangle$   &$P_{01}(1810)$ &                    &$P_{03}(?)$\\
 $|70,^28,2,0\rangle$    &$P_{01}(?)$ &                    &$F_{05}(?)$\\
$|70,^48,2,0\rangle$    &$P_{03}(?)$   &                    &$F_{07}(?)$\\
\hline
\end{tabular}
\end{table}

\begin{table}[ht]
\caption{Breit-Wigner masses $M_R$ (MeV) and widths $\Gamma_R$ (MeV)
for the resonances in the $s$-channel compared with the experimental
data from PDG.} \label{BW}
\begin{tabular}{|c|c|c|c|c|c|c|c|c|c|c }\hline\hline
Resonance &$M_R$  &$\Gamma_R$   &$M_R$ (PDG) &$\Gamma_R$ (PDG)\\
\hline
$\Lambda(1405)S_{01}$ &$1405.0$ &$53.37$ &$1405.0^{+1.3}_{-1.0}$ &$50\pm2$\\
$\Lambda(1520)D_{03}$ &$1519.5$ &$15.6$ &$1519.5\pm1.0$&$15.6\pm1.0$\\
\hline
$\Lambda(1670)S_{01}$ &$1676.0$&$35.0$&$1670\pm10$&$25\sim50$\\
$\Lambda(1690)D_{03}$  &$1682.4$ &$70.0$ &$1690\pm5$&$50\sim70$\\
\hline
\end{tabular}
\end{table}

\subsection{Total cross section}

\begin{figure}[ht]
\centering \epsfxsize=9.0 cm
\epsfbox{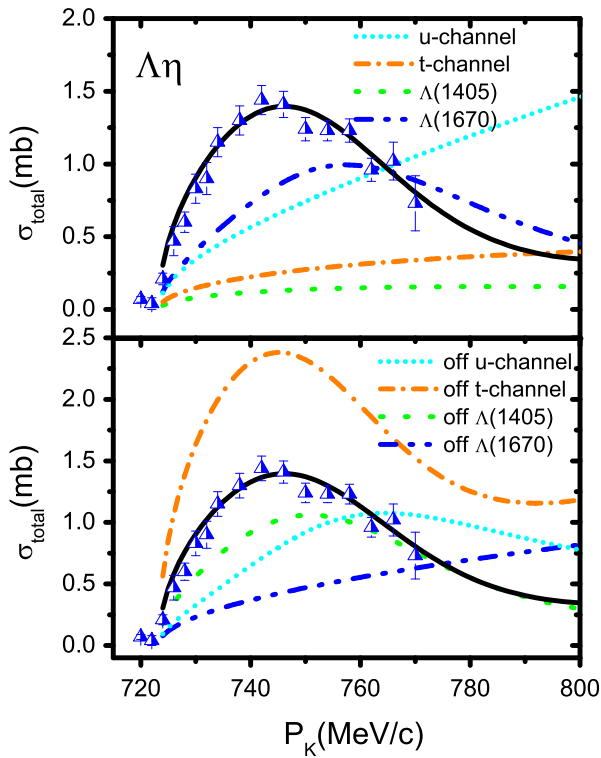}\caption{$K^-p\rightarrow\Lambda\eta$ total cross
sections compared with the data~\cite{Starostin:2001zz}. The bold
solid curves are for the full model calculations. In the upper
panel, exclusive cross sections for $\Lambda(1405)S_{01}$,
$\Lambda(1670)S_{01}$,  $t$-channel and $u$-channel are indicated
explicitly by the legends in the figures. In the lower panel, the
results by switching off the contributions of $\Lambda(1405)S_{01}$,
$\Lambda(1670)S_{01}$, $t$-channel and $u$-channel are indicated
explicitly by the legends in the figures.} \label{fig-tcr}
\end{figure}

The total cross section as a function of the beam momentum is shown
in Fig.~\ref{fig-tcr}, where we find that the observations can be
well described within the chiral quark model.

It is found that $\Lambda(1670)S_{01}$ should play a dominant role
in the reaction. $\Lambda(1670)S_{01}$ is responsible for the bump
structure in the cross section around its threshold. To well
describe the data, a large amplitude of $\Lambda(1670)S_{01}$ in the
reaction is needed, which is about 34 times (i.e.,
$C_{S_{01}(1670)}=34 $) larger than that derived in the
$SU(6)\otimes O(3)$ limit. In our previous work, we found
$\Lambda(1670)S_{01}$ should have a weaker coupling to $\bar{K}N$
than that derived in the $SU(6)\otimes O(3)$
limit~\cite{Zhong:2013oqa}, thus, $\Lambda(1670)S_{01}$ should have
a much stronger coupling to $\eta \Lambda$ than that derived from
the symmetry quark model. These phenomenologies might be explained
by the configuration mixing between the $S$-wave states
$\Lambda(1405)S_{01}$, $\Lambda(1670)S_{01}$ and
$\Lambda(1800)S_{01}$, which will be further studied in
Sec.~\ref{cxx}.

Furthermore, a sizeable contribution from $\Lambda(1405)$ also can
be seen in the cross section. The total cross section around the
peak is slightly underestimated without its contribution.

No obvious contributions from the $D$-wave states,
$\Lambda(1520)D_{03}$ and $\Lambda(1690)D_{03}$, are found in the
reaction.

It should be emphasized that both $u$- and $t$-channel backgrounds
play crucial roles in the reactions. Switching off their
contributions, the cross section changes significantly. The
important roles of $u$- and/or $t$-channel backgrounds are also
found in the other $\bar{K}N$ reactions $K^-p\rightarrow
\Sigma^0\pi^0,\Lambda\pi^0,\bar{K}^0n$~\cite{Zhong:2013oqa}.

\begin{figure*}[ht]
\centering \epsfxsize=16.0 cm \epsfbox{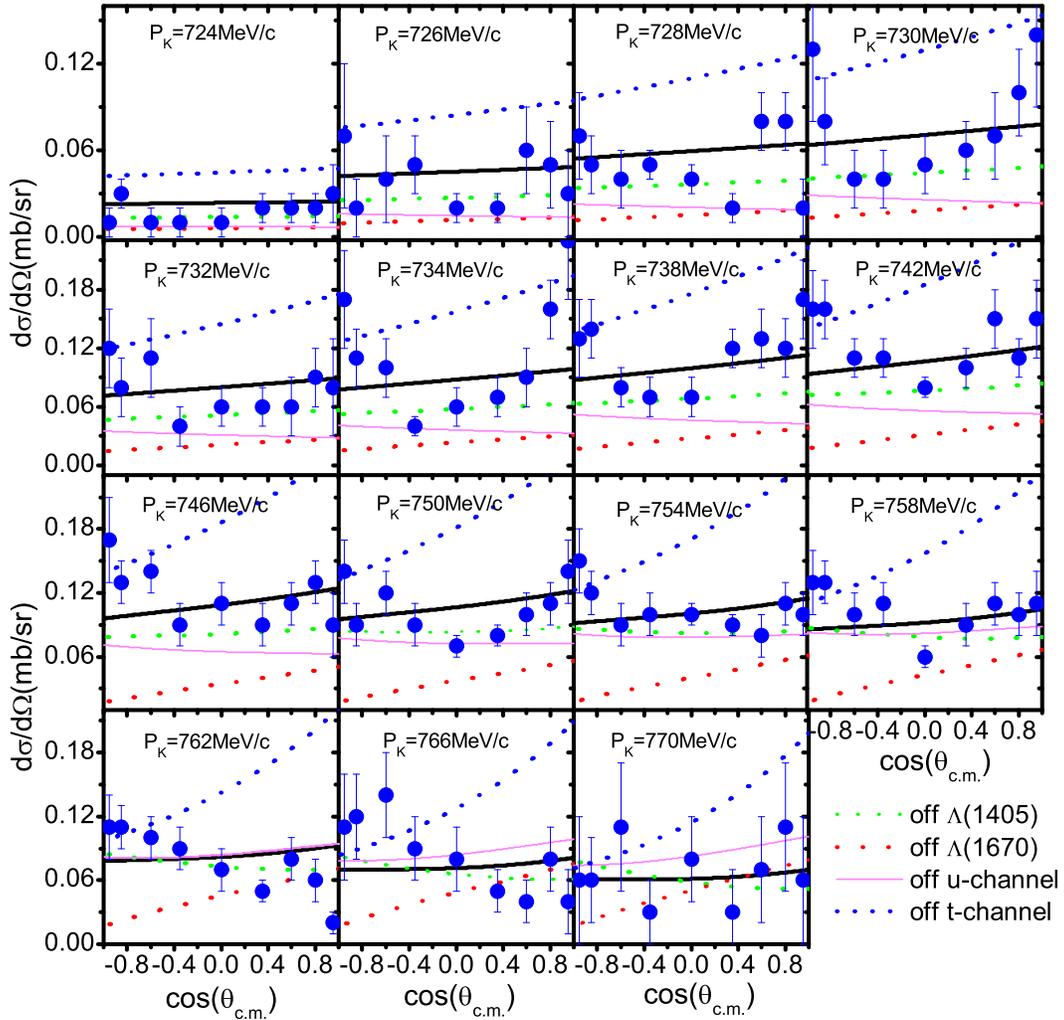}
\caption{Differential cross sections of the $K^-p\rightarrow
\eta\Lambda$ compared with the data from~\cite{Starostin:2001zz}.
The bold solid curves are for the full model calculations. The
results by switching off the contributions from
$\Lambda(1405)S_{01}$, $\Lambda(1670)S_{01}$ and $u$- and
$t$-channel backgrounds are indicated explicitly by the legend in
the figures. }\label{fig-sd}
\end{figure*}

\subsection{Differential cross section}

The differential cross sections (DCS) compared with the data are
shown in Fig.~\ref{fig-sd}. From the figure, it is seen that the
data in the low energy region from threshold to $P_K=770$ MeV/c can
be reasonably described within our chiral quark model. However, it
should be remarked that our theoretical results seem to slightly
underestimate the DCS at both forward and backward angles in the
beam momenta region of $P_K=730\sim742$ MeV/c. Improved measurements
in this beam momenta region are needed to clarify the discrepancies.

To explore the contribution of individual resonances and the $u$-
and $t$-channel backgrounds to the DCS, we have shown the
predictions by switching off one of their contributions in
Fig.~\ref{fig-sd} as well. From the figure, the dominant roles of
$\Lambda(1670)S_{01}$ and $u$-, $t$-channel backgrounds are
significantly seen in the DCS. Switching off the contribution of
$\Lambda(1670)S_{01}$, we find that the cross sections will be
underestimated draftily. Without the $u$-channel contribution, the
DCS will be significantly underestimated around $\eta$ production
threshold. While, switching off $t$-channel contribution, we can see
that the DCS are strongly overestimated at both forward and backward
angles. Furthermore, slight contributions of $\Lambda(1405)S_{01}$
can be seen in the DCS around $\eta$ production threshold, where
$\Lambda(1405)S_{01}$ has a constructive interference with
$\Lambda(1670)S_{01}$. However, $\Lambda(1405)S_{01}$ is not a
crucial contributor in the reaction, which can explain the reason
why the contribution of $\Lambda(1405)S_{01}$ can be neglected in
some studies at the hadron
level~\cite{Manley:2002,Liu:2012ge,Liu:2011sw}.

From Fig.~\ref{fig-sd}, it is seen that a bowl structure seems to
appear in the data around $\eta$ production threshold. As we know,
the bowl structures are the typical effects of the interferences
between the $S$- and $D$-wave states. In this energy region, the
bowl structures might be caused by the interferences between
$\Lambda(1670)S_{01}$ and $\Lambda(1690)D_{03}$. Considering
$\Lambda(1690)D_{03}$ as the conventional three-quark state
classified in the constituent quark model, we can not obtain a bowl
structure in the DCS for the too small contributions of
$\Lambda(1690)D_{03}$ in the reaction. In
Refs.~\cite{Liu:2012ge,Liu:2011sw}, Liu and Xie had carefully
studied these bowl structures appearing in the DCS, they need an
exotic $D$-wave state $\Lambda(1669)$ with a very narrow width of
$\Gamma\simeq 1.5$ MeV to reproduce the bowl structures. Finally, it
should be pointed out that for the rather large uncertainties of the
present data, we can not confirm whether there are obvious bowl
structures in the DCS or not. Thus, more accurate measurements are
needed.

As a whole, $\Lambda(1670)S_{01}$ plays a dominant role in the
reaction. $\Lambda(1670)S_{01}$ should have a much stronger coupling
to $\eta \Lambda$, while have a weaker coupling to $\bar{K}N$ than
that derived in the $SU(6)\otimes O(3)$ limit. The $u$- and
$t$-channel backgrounds also play crucial roles in the reaction.
Furthermore, slight contributions of $\Lambda(1405)S_{01}$ can be
seen in the DCS around $\eta$ production threshold. No obvious
evidence from the $D$-wave states, $\Lambda(1520)D_{03}$ and
$\Lambda(1690)D_{03}$, is found in the reaction.

\subsection{Configuration mixing and strong couplings}\label{cxx}

To further understand the strength parameters $C_R$ in the
$K^-p\rightarrow\Lambda\eta$ reaction, and explain the strong
coupling properties of the $\Lambda$ resonances extracted from the
$\bar{K}N$ scattering, e.g., the weak coupling of
$\Lambda(1670)S_{01}$ to $\bar{K}N$ and strong coupling of
$\Lambda(1670)S_{01}$ to $\eta \Lambda$, in this subsection we study
the configuration mixing effects in the low-lying negative $\Lambda$
resonances.

\subsubsection{Configuration mixing and strong decays}

If there is configuration mixing in several resonances with the same
$J^P$ values, their strong coupling properties might be very
different from the pure states classified in the constituent quark
model. Here, we study the strong decays of low-lying negative
$\Lambda$ resonances and test whether the configuration mixing can
explain the strong couplings of these resonances or not.

In this work, the strong decays of the $\Lambda$ resonances also
studied with the chiral quark model. This approach has been
successfully used to study the strong decays of charmed baryons,
$\Xi$ baryons, and heavy-light
mesons~\cite{Zhong:2007gp,Zhong:2010vq,Liu:2012sj,Xiao:2013xi}. The
details of how to describe the strong decays of the baryon
resonances in the chiral quark model can be found
in~\cite{Xiao:2013xi}.

As we know, $\Sigma(1385)$ is a well-estimated strangeness-1 hyperon
state. According to the classification of the quark model, it is
assigned to the pure $|56,^410,0,0,\frac{3}{2}^+\rangle$
representation. In this work, the measured width of $\Sigma^0(1385)$
as an input (i.e., $\Gamma=36$ MeV~\cite{Beringer:2012}) is used to
determine the overall parameter $\delta$($=0.654$) in the decay
amplitudes. With this determined parameter, we can calculate the
strong decays of the other strangeness-1 hyperon states.

\paragraph{$S$-wave states}

Firstly, we study the strong decay properties of the $S$-wave states
$\Lambda(1405)S_{01}$, $\Lambda(1670)S_{01}$ and
$\Lambda(1800)S_{01}$. If they are pure states, according to the
classification of the constituent quark model, they should be
assigned to the $|70,^21,1,1,\frac{1}{2}^-\rangle$,
$|70,^28,1,1,\frac{1}{2}^-\rangle$ and
$|70,^48,1,1,\frac{1}{2}^-\rangle$, respectively~\cite{Xiao:2013xi}.

\begin{table}[ht]
\caption{The predicted total and partial decay widths (MeV) and
partial decay width ratios of $\Lambda(1670)S_{01}$ as a pure state
of $|70,^28,1,1,\frac{1}{2}^-\rangle$. $\Gamma^{th}$ denotes our
prediction, while $\Gamma^{exp}$ denotes the data from PDG. }
\label{lambda16}
\begin{tabular}{|c|c|c|c|c|c|c|c|c|c|c }\hline\hline
Channel&$\Gamma^{th}_i$&$\Gamma^{th}_{total}$&$\Gamma^{exp}_{total}$&$
\frac{\Gamma_i}{\Gamma_{total}}|_{th}$~~~~~&$\frac{\Gamma_i}{\Gamma_{total}}|_{exp}$\\
\hline
$\Sigma\pi$~~~~~&$15.4$&$123.4$&$25$ to $50(\approx35)$&$0.12$&$0.25\sim0.55$\\
$NK$~~~~~&$103.1$~~~~~&$$~~~~~&$$~~~~~&$0.84$&$0.20\sim0.30$\\
$\Lambda\eta$~~~~~&$0.28$~~~~~&$$~~~~~&$$~~~~~&$0.002$&$0.10\sim0.25$\\
$\Sigma(1385)\pi$~~~~~&$4.7$~~~~~&$$~~~~~&$$~~~~~&$0.04$&$\cdot\cdot\cdot$\\
\hline\hline
\end{tabular}
\end{table}

Considering $\Lambda(1670)S_{01}$ as the pure state
$|70,^28,1,1,\frac{1}{2}^-\rangle$, we calculate its strong decay
properties, which are listed in Tab.~\ref{lambda16}. From the table,
we see that the total decay width in theory
($\Gamma^{th}_{total}=123.4$ MeV) are much larger than the data
($\Gamma^{exp}_{total}\simeq35$ MeV). Meanwhile, according to our
calculation, the branching ratio of the $\Lambda\eta$ channel is too
small, while the branching ratio of the $N\bar{K}$ channel is too
large to compare with the data. Thus, as a pure state, the
$\Lambda(1670)S_{01}$ strong decays can not be described at all.

It should be remarked that several different representations with
the same $J^P$ numbers might be coupled together via some kind of
interactions~\cite{Isgur78,Isgur:1977ky,Capstick:1986bm,Schat:2001xr}.
Thus, $\Lambda(1670)S_{01}$ may be a mixed state between three
different representations $|70,^21,1,1\rangle$, $|70,^28,1,1\rangle$
and $|70,^48,1,1\rangle$ with $J^P=1/2^-$. Based on the standard
Cabibbo-Kobayashi-Maskawa (CKM) matrix method, the physical states
might be expressed as
\begin{equation}\label{mixs}
\left(\begin{array}{c}|\Lambda(1800)\frac{1}{2}^-\rangle\cr
|\Lambda(1670)\frac{1}{2}^-\rangle\cr
 |\Lambda(1405)\frac{1}{2}^-\rangle
\end{array}\right)=U \left(\begin{array}{c} |70,^21\rangle \cr |70,^28\rangle \cr |70,^48\rangle
\end{array}\right),
\end{equation}
with
\begin{equation}
U=\left(\begin{array}{ccc} c_{12}c_{13} & s_{12}c_{13} & s_{13} \cr
-s_{12}c_{23}-c_{12}s_{23}s_{13} & c_{12}c_{23}-s_{12}s_{23}s_{13} &
s_{23}c_{13} \cr s_{12}s_{23}-c_{12}c_{23}s_{13} &
-c_{12}s_{23}-s_{12}c_{23}s_{13} & c_{23}c_{13}
\end{array}\right),
\end{equation}
where $c_{ij}\equiv \cos{\theta}_{ij}$ and $s_{ij}\equiv
\sin{\theta}_{ij}$. $\theta_{ij}$ stands for the mixing angles,
which could be determined by fitting the strong decay data of
$\Lambda(1670)S_{01}$.

By fitting the experiment data from PDG~\cite{Beringer:2012}, we
have obtained that $\theta_{12}\simeq 75^0$, $\theta_{13}\simeq
50^0$ and $\theta_{23}\simeq 125^0$. With these mixing angles, the
strong decay properties of $\Lambda(1670)S_{01}$ can be reasonably
described. The theoretical results compared with the data are listed
in Tab.~\ref{mix16}. From the table, it is seen that with the
configuration mixing the $\Lambda\eta$ branching ratio is enhanced
obviously, while the $N\bar{K}$ branching ratio is suppressed, which
can naturally explain the weak coupling of $\Lambda(1670)S_{01}$ to
$\bar{K}N$ and strong coupling of $\Lambda(1670)S_{01}$ to $\eta
\Lambda$ needed in the $\bar{K}N$ reactions.

\begin{table}[ht]
\caption{The predicted total and partial decay widths (MeV) and
partial decay width ratios of $\Lambda(1670)S_{01}$ as a mixed state
compared with the experiment data from PDG.} \label{mix16}
\begin{tabular}{|c|c|c|c|c|c|c|c|c|c|c }\hline\hline
Channel&$\Gamma^{th}_i$~~~~~&$\Gamma^{th}_{total}$~~~~~&$\Gamma^{exp}_{total}$&$\frac{\Gamma_i}{\Gamma_{total}}|_{th}$~~~~~&$\frac{\Gamma_i}{\Gamma_{total}}|_{exp}$\\
\hline
$\Sigma\pi$~~~~~&$11.8$~~~~~&$44.7$~~~~~&$25$ to $50(\approx35)$&$0.26$~~~~~&$0.25\sim0.55$\\
$NK$~~~~~&$13.6$~~~~~&$$~~~~~&$$~~~~~&$0.30$~~~~~&$0.20\sim0.30$\\
$\Lambda\eta$~~~~~&$18.2$~~~~~&$$~~~~~&$$~~~~~&$0.41$~~~~~&$0.10\sim0.25$\\
$\Sigma(1385)\pi$~~~~~&$1.1$~~~~~&$$~~~~~&$$~~~~~&$0.02$~~~~~&$\cdot\cdot\cdot$\\
\hline\hline
\end{tabular}
\end{table}

According to the determined mixing angles, Eq.(\ref{mixs}) can be
explicitly expressed as
\begin{equation}\label{mix1}
\left(\begin{array}{c}|\Lambda(1800)\frac{1}{2}^-\rangle\cr |\Lambda(1670)\frac{1}{2}^-\rangle\cr
 |\Lambda(1405)\frac{1}{2}^-\rangle\cr\end{array}\right)=\left(\begin{array}{ccc}
0.17&0.62&0.77\cr 0.39 &-0.76&0.53\cr 0.90&0.21&-0.37
\end{array}\right)\left(\begin{array}{c}|70,^21\rangle\cr|70,^28\rangle\cr|70,^48\rangle\cr
\end{array}\right),
\end{equation}
where, we find that the main component of $\Lambda(1670)S_{01}$ is
$|70,^28\rangle (\sim58\%$). Meanwhile, the $|70,^21\rangle$ and
$|70,^48\rangle$ components also have a sizable proportion, which
are $\sim 15\%$ and $\sim 28\%$, respectively. $\Lambda(1405)S_{01}$
is dominated by the $|70,^21\rangle$($\sim81\%$), while it contains
significant octet components of $|70,^28\rangle(\sim4\%$) and
$|70,^48\rangle$($\sim14\%$). $\Lambda(1800)S_{01}$ is dominated by
both the $|70,^48\rangle$($\sim59\%$) and $|70,^28\rangle(\sim38\%$)
components.

With these mixing schemes, we have calculated the strong decay
properties of $\Lambda(1405)S_{01}$ and $\Lambda(1800)S_{01}$. The
calculated decay width of $\Lambda(1405)S_{01}$ is $\Gamma\simeq53$
MeV, which in good agreement with the data ($\Gamma=50\pm2$ MeV).

\begin{figure}[ht]
\centering \epsfxsize=9.0 cm \epsfbox{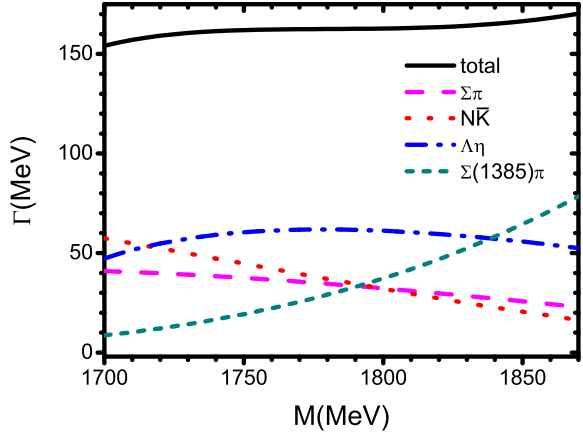} \caption{The strong
decay properties of $\Lambda(1800)S_{01}$, which is taken as a mixed
state in Eq.(\ref{mixs}).}\label{fig-s8}
\end{figure}

Considering the uncertainties in the mass of $\Lambda(1800)S_{01}$,
we vary its mass from 1700 MeV to 1870 MeV. The predicted strong
decay properties of $\Lambda(1800)S_{01}$ have been shown in
Fig.~\ref{fig-s8}. From the figure, we can see that the strong
decays of $\Lambda(1800)S_{01}$ are dominated by the $\bar{K}N$,
$\eta \Lambda$ and $\Sigma \pi$ decay modes. While the decay channel
$\Sigma(1385)\pi$ also has a significant contribution to the strong
decays of $\Lambda(1800)S_{01}$. It is found that our predicted
strong decay properties of $\Lambda(1800)S_{01}$ are compatible with
the data of ALSTON (see
Tab.~\ref{w180})~\cite{AlstonGarnjost:1977rs}. About
$\Lambda(1800)S_{01}$, more measurements are needed in experiments.

As a whole the configuration mixing is needed to understand the
strong decay properties of the $S$-wave resonances
$\Lambda(1405)S_{01}$, $\Lambda(1670)S_{01}$ and
$\Lambda(1800)S_{01}$.

\begin{table}[ht]
\caption{The predicted total and partial decay widths (MeV) of
$\Lambda(1800)S_{01}$ compared with the experiment data from
ALSTON~\cite{AlstonGarnjost:1977rs}. We set the mass of
$\Lambda(1800)S_{01}$ as M=1725 MeV, which is the observed value
from ALSTON.} \label{w180}
\begin{tabular}{|c|c|c|c|c|c|c|c|c|c|c }\hline\hline
Channel&$N\bar{K}$~~~~~&$\Sigma\pi$~~~~~&$\Lambda\eta$~~~~~&$\Sigma(1385)\pi$~~~~~\\
\hline
$\Gamma^{th}_i$&$51.1$&$39.5$&$56.1$&$13.2$ \\
\hline
$\Gamma^{exp}_{i}$&$52\pm 9$&...&...~ &... \\
\hline\hline
\end{tabular}
\end{table}

\paragraph{$D$-wave states}

Then, we will further study whether the configuration mixing is
necessary to explain the strong decays of the well-established
$D$-wave resonances $\Lambda(1520)D_{03}$ and $\Lambda(1690)D_{03}$
or not. If $\Lambda(1520)D_{03}$ and $\Lambda(1690)D_{03}$ are pure
states, they should be classified as the
$|70,^21,1,1,\frac{3}{2}^-\rangle$ and
$|70,^48,1,1,\frac{3}{2}^-\rangle$ configurations, respectively, in
the constituent quark model.

\begin{table}[ht]
\caption{The predicted total and partial decay widths (MeV) and
partial decay width ratios of $\Lambda(1520)D_{03}$ as a pure state
$|70,^21,1,1,\frac{3}{2}^-\rangle$ compared with the experiment data
from PRD.} \label{w150}
\begin{tabular}{|c|c|c|c|c|c|c|c|c|c|c }\hline\hline
Channel&$\Gamma^{th}_i$~~~~~&$\Gamma^{th}_{total}$~~~~~&$\Gamma^{exp}_{total}$
~~~~~&$\frac{\Gamma_i}{\Gamma_{total}}|_{th}$~~~~~&$\frac{\Gamma_i}{\Gamma_{total}}
|_{exp}$\\
\hline
$\Sigma\pi$~~~~~&$10.7$~~~~~&$14.5$~~~~~&$15.6\pm1.0$~~~~~&$0.74$~~~~~&$0.42\pm0.01$\\
$NK$~~~~~&$3.81$~~~~~&$$~~~~~&$$~~~~~&$0.26$~~~~~&$0.45\pm0.01$\\
\hline\hline
\end{tabular}
\end{table}

\begin{table}[ht]
\caption{The predicted total and partial decay widths (MeV) and
partial decay width ratios of $\Lambda(1690)D_{03}$ as a pure state
of $|70,^28,1,1,\frac{3}{2}^-\rangle$ compared with the experiment
data from PRD.} \label{w1690}
\begin{tabular}{|c|c|c|c|c|c|c|c|c|c|c }\hline\hline
Channel&$\Gamma^{th}_i$&$\Gamma^{th}_{total}$&$\Gamma^{exp}_{total}$&$
\frac{\Gamma_i}{\Gamma_{total}}|_{th}$&$\frac{\Gamma_i}{\Gamma_{total}}|_{exp}$\\
\hline $\Sigma\pi$&$9.74$&$117.2$&$50\sim70(\approx60)$&$0.08$&$
0.20\sim0.40$\\
$NK$&$58.31$&$$&$$&$0.50$&$0.20\sim0.30$\\
$\Lambda\eta$&$0.001$~~~~~&$$~~~~~&$$~~~~~&$0.00$&$\cdot\cdot\cdot$\\
$\Sigma(1385)\pi$~~~~~&$49.1$~~~~~&$$~~~~~&$$~~~~~&$0.42$&$\cdot\cdot\cdot$\\
\hline\hline
\end{tabular}
\end{table}

Firstly, we study the decay properties of $\Lambda(1520)D_{03}$ and
$\Lambda(1690)D_{03}$ as pure states. The predictions compared with
the data are listed in Tabs.~\ref{w150} and ~\ref{w1690},
respectively.

From Tab.~\ref{w150}, we find that as a pure state the strong decay
coupling of $\Lambda(1520)D_{03}$ to $\Sigma\pi$ is overestimated.
However, the strong coupling of $\Lambda(1520)D_{03}$ to $N\bar{K}$
is underestimated, which is also found in the $\bar{K}N$
scattering~\cite{Zhong:2013oqa}.

While considering $\Lambda(1690)D_{03}$ as a pure state
$|70,^48,1,1,\frac{3}{2}^-\rangle$, from Tab.~\ref{w1690} we find
that the theoretical predictions are inconsistent with the
experimental observations. The predicted total decay width is much
larger than the data. In addition, the partial decay width ratio for
$\Sigma\pi$ is too small, while, that for $N\bar{K}$ is too large to
compare with the data. Thus, as pure states, the strong decay
properties of $\Lambda(1520)D_{03}$ and $\Lambda(1690)D_{03}$ can't
be understood reasonably.

For these reasons, it is naturally for us to take $\Lambda(1520)$
and $\Lambda(1690)$ as two mixing states between
$|70,^21,1,1,\frac{3}{2}^-\rangle$,
$|70,^28,1,1,\frac{3}{2}^-\rangle$ and
$|70,^48,1,1,\frac{3}{2}^-\rangle$. By the using of the CKM matrix
method again, and fitting the strong decay data of $\Lambda(1690)$,
we obtain
\begin{equation}\label{mix2}
\left(\begin{array}{c}|\Lambda(1520)\frac{3}{2}^-\rangle\cr |\Lambda(1690)\frac{3}{2}^-\rangle\cr
 |\Lambda\frac{3}{2}^-\rangle_3\cr\end{array}\right)=\left(\begin{array}{ccc}
0.94&0.34&0.09\cr
0.31&-0.92&0.26\cr
0.17&-0.21&-0.96
\end{array}\right)\left(\begin{array}{c}|70,^21\rangle\cr|70,^28\rangle\cr|70,^48\rangle\cr
\end{array}\right).
\end{equation}

From Eq.(\ref{mix2}), it is seen that $\Lambda(1690)$ has sizable
components of $|70,^21\rangle$ ($\sim 9\%$) and $|70,^48\rangle$
($\sim 7\%$), except for the main component $|70,^28\rangle$ ($\sim
85\%$).  The predicted strong decay properties of
$\Lambda(1690)D_{03}$ compared with the data are listed in
Tab.~\ref{mix80}, where we find that with the configuration mixing
effects, the strong decays of $\Lambda(1690)D_{03}$ can be well
described. It should be emphasized that $\Lambda(1690)$ has a very
weak coupling to $\Lambda\eta$, although it has been draftily
enhanced by considering the configuration mixing effects, which
gives an explanation why the contribution of $\Lambda(1690)D_{03}$
to the reaction $K^-P\rightarrow\Lambda\eta$ is tiny even though
$\Lambda(1690)D_{03}$ has a large $C_R$-factor.

\begin{table}[ht]
\caption{The predicted total and partial decay widths (MeV) and
partial decay width ratios of $\Lambda(1520)$ as a mixed state
compared with the experiment data from PDG.} \label{m150}
\begin{tabular}{|c|c|c|c|c|c|c|c|c|c|c }\hline\hline
Channel&$\Gamma^{th}_i$~~~~~&$\Gamma^{th}_{total}$~~~~~&$\Gamma^{exp}_{total}$~~~~~
&$\frac{\Gamma_i}{\Gamma_{total}}|_{th}$&$\frac{\Gamma_i}{\Gamma_{total}}|_{exp}$\\
\hline
$\Sigma\pi$~~~~~&$7.0$~~~~~&$13.5$~~~~~&$15.6\sim1.0$~~~~~&$0.52$&$0.42\pm0.01$\\
$NK$~~~~~&$6.2$~~~~~&$$~~~~~&$$~~~~~&$0.46$&$0.45\pm0.01$\\
$\Sigma(1385)\pi$~~~~~&$0.3$~~~~~&$$~~~~~&$$~~~~~&$0.02$&$\cdot\cdot\cdot$\\
\hline\hline
\end{tabular}
\end{table}

\begin{figure}[ht]
\centering \epsfxsize=9.0 cm \epsfbox{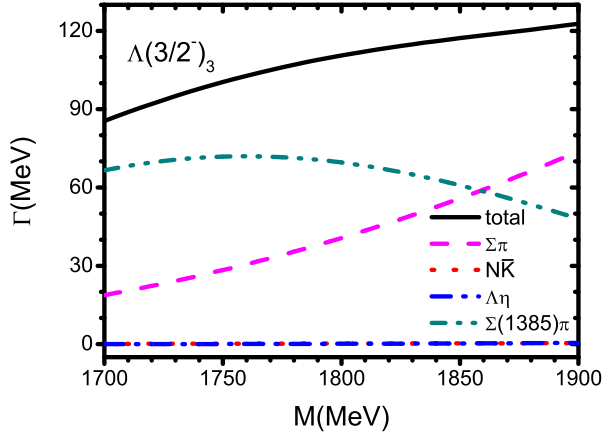}
\caption{The strong decay properties of
$|\Lambda\frac{3}{2}^-\rangle_3$ as a counterpart of
$\Lambda(1690)$.}\label{fig-dw}
\end{figure}

Furthermore, with the mixing scheme determined in Eq.(\ref{mix2}),
we study the strong decay of $\Lambda(1520)D_{03}$. The predicted
results compared with the data are listed in Tab.~\ref{m150}, where
we find both the total decay width and the partial decay width
ratios are in good agreement with the data. The $N\bar{K}$ branching
ratio is about a factor 2 larger than that derived in the
$SU(6)\otimes O(3)$ limit, which is consistent with our previous
analysis of the $\bar{K}N$ scattering in~\cite{Zhong:2013oqa}. From
Eq.(\ref{mix2}), we can see that the main component of
$\Lambda(1520)D_{03}$ is still the $|70,^21\rangle$ configuration
($\sim 88\%$), while it contains significant octet component of
$|70,^28\rangle$ ($\sim 12\%$).

\begin{table}[ht]
\caption{The predicted total and partial decay widths(MeV) and
partial decay width ratios of $\Lambda(1690$ as a mixed state
compared with the experiment data from PDG. }%The mixing angles are
%$\theta_{12}=20^0$, $\theta=5^0$ and $\theta_{23}=165^0$}
\label{mix80}
\begin{tabular}{|c|c|c|c|c|c|c|c|c|c|c }\hline\hline
Channel&$\Gamma^{th}_i$~~~~~&$\Gamma^{th}_{total}$&$\Gamma^{exp}_{total}$
&$\frac{\Gamma_i}{\Gamma_{total}}|_{th}$&$\frac{\Gamma_i}{\Gamma_{total}}|_{exp}$\\
\hline
$\Sigma\pi$&$27.5$&$70.6$&$50\sim70(\approx60)$&$0.39$&$
0.20\sim0.40$\\
$NK$&$21.4$&$$&$$&$0.30$&$0.20\sim0.30$\\
$\Lambda\eta$&$0.01$&$$&$$&$0.00$&$\cdot\cdot\cdot$\\
$\Sigma(1385)\pi$&$21.6$&$$&$$&$0.30$&$\cdot\cdot\cdot$\\
\hline\hline
\end{tabular}
\end{table}

Finally, we give our predictions of the third $D$-wave resonance
$|\Lambda\frac{3}{2}^-\rangle_3$, which is still not established in
experiment. According to the quark model prediction, its mass is
around $1800$
MeV~\cite{Isgur78,Isgur:1977ky,Capstick:1986bm,Schat:2001xr}.
Varying its mass from 1700 MeV to 1900 MeV, we calculate the strong
decays of $|\Lambda\frac{3}{2}^-\rangle_3$. Our predictions have
been shown in Fig.~\ref{fig-dw}. It is found that the strong decays
of $|\Lambda\frac{3}{2}^-\rangle_3$ are dominated by
$\Sigma(1385)\pi$ and $\Sigma\pi$, while the $N\bar{K}$ and $\Lambda
\eta$ branching ratios are negligibly small. Thus, we suggest the
our experimental colleagues find this missing $D$-wave state in the
$\Sigma(1385)\pi$ and $\Sigma\pi$ channels.

In a word, the configuration mixing is also needed to understand the
strong decay properties of the $D$-wave resonances
$\Lambda(1520)D_{03}$, $\Lambda(1690)D_{03}$.

\subsubsection{Interpretation of $C_R$ with configuration mixing}

If the configuration mixing effects are included, the
single-resonance-excitation amplitude given in Eq.~(\ref{pt}) should
be rewritten as
\begin{eqnarray}
\mathcal{O}(n,l,J)=\sum_Rg_R'\mathcal{O}(n,l,J)\equiv\sum_RC_Rg_R\mathcal{O}(n,l,J),
\end{eqnarray}
where $g_R'$ ($g_R$) stands for the relative strength of a
single-resonance with (without) configuration mixing effects in the
partial amplitude $\mathcal{O}(n,l,J)$. The $C_R$ parameters can be
derived by
\begin{equation}\label{cp}
C_R=\frac{g'_R}{g_R}.
\end{equation}
In the following work, we study the $C_R$ parameters of the
important resonances $\Lambda(1405)S_{01}$, $\Lambda(1670)S_{01}$,
$\Lambda(1520)D_{03}$ and $\Lambda(1690)D_{03}$ for the
$K^-p\rightarrow \eta\Lambda$ process.

Taking $\Lambda(1405)S_{01}$, $\Lambda(1670)S_{01}$,
$\Lambda(1520)D_{03}$ and $\Lambda(1690)D_{03}$ as pure states in
the constituent quark model, we can derive the couplings of the
transition amplitudes for these resonances, which are given by
\begin{eqnarray}
R_{\Lambda(1405)}&=&-\frac{\sqrt{3}}{108}(\sqrt{2}\sin\phi_P+\cos\phi_P),\\
R_{\Lambda(1670)}&=&-\frac{\sqrt{3}}{108}(\sqrt{2}\sin\phi_P-\cos\phi_P),\\
R_{\Lambda(1520)}&=&-\frac{\sqrt{3}}{54}(\sqrt{2}\sin\phi_P+\cos\phi_P),\\
R_{\Lambda(1690)}&=&-\frac{\sqrt{3}}{54}(\sqrt{2}\sin\phi_P-\cos\phi_P),
\end{eqnarray}
where the $\phi_P$ is the $\eta$-$\eta'$ mixing angle. Then the
$g_R$ parameters for these states can be obtained by
\begin{eqnarray}
g_{\Lambda(1405)~ \mathrm{or}~
\Lambda(1670)}&=&\frac{R_{\Lambda(1405)}~
\mathrm{or}~R_{\Lambda(1670)}}{R_{\Lambda(1405)}+R_{\Lambda(1670)}},\\
g_{\Lambda(1520)~ \mathrm{or}~
\Lambda(1690)}&=&\frac{R_{\Lambda(1520)}~
\mathrm{or}~R_{\Lambda(1690)}}{R_{\Lambda(1520)}+R_{\Lambda(1690)}}.
\end{eqnarray}

Considering the configuration mixing effects, the wave functions of
the $S$- and $D$-wave states $\Lambda(1405)S_{01}$,
$\Lambda(1670)S_{01}$, $\Lambda(1520)D_{03}$ and
$\Lambda(1690)D_{03}$ can be generally written as
\begin{eqnarray}
|\Lambda(1405)\rangle=a_1 |70,^21\rangle_{S}+b_1
|70,^28\rangle_{S}+c_1|70,^48\rangle_{S},\\
|\Lambda(1670)\rangle=a_2 |70,^21\rangle_S+b_2 |70,^28\rangle_S+c_2
|70,^48\rangle_S,\\
|\Lambda(1520)\rangle=a_3 |70,^21\rangle_{D}+b_3
|70,^28\rangle_{D}+c_3|70,^48\rangle_{D},\\
|\Lambda(1690)\rangle=a_4 |70,^21\rangle_D+b_4 |70,^28\rangle_D+c_4
|70,^48\rangle_D,
\end{eqnarray}
where $a_i$, $b_i$ and $c_i$ ($i=1,2,3,4$) have been determined in
Eqs.~(\ref{mix1}) and (\ref{mix2}). Then we can derive the couplings
of the transition amplitudes for these mixed states, they are
\begin{eqnarray}
R'_{\Lambda(1405)}&=&-\frac{\sqrt{3}}{108}(\sqrt{2}\sin\phi_P+\cos\phi_P)(a_1^2+a_1b_1)\nonumber\\
&&-\frac{\sqrt{3}}{108}(\sqrt{2}\sin\phi_P-\cos\phi_P)(b_1^2+a_1b_1),\\
R'_{\Lambda(1670)}&=&-\frac{\sqrt{3}}{108}(\sqrt{2}\sin\phi_P+\cos\phi_P)(a_2^2+a_2b_2)\nonumber\\
&&-\frac{\sqrt{3}}{108}(\sqrt{2}\sin\phi_P-\cos\phi_P)(b_2^2+a_2b_2),\\
R'_{\Lambda(1520)}&=&-\frac{\sqrt{3}}{54}(\sqrt{2}\sin\phi_P+\cos\phi_P)(a_3^2+a_3b_3)\nonumber\\
&&-\frac{\sqrt{3}}{54}(\sqrt{2}\sin\phi_P-\cos\phi_P)(b_3^2+a_3b_3),\\
R'_{\Lambda(1690)}&=&-\frac{\sqrt{3}}{54}(\sqrt{2}\sin\phi_P+\cos\phi_P)(a_4^2+a_4b_4)\nonumber\\
&&-\frac{\sqrt{3}}{54}(\sqrt{2}\sin\phi_P-\cos\phi_P)(b_4^2+a_4b_4).
\end{eqnarray}
Finally, we obtain the relative strength parameters $g_R'$ for these
mixed states:
\begin{eqnarray}
g_{\Lambda(1405)~ \mathrm{or}~
\Lambda(1670)}'&=&\frac{R_{\Lambda(1405)}'~
\mathrm{or}~R_{\Lambda(1670)}'}{R_{\Lambda(1405)}'+R_{\Lambda(1670)}'},\\
g_{\Lambda(1520)~ \mathrm{or}~
\Lambda(1690)}'&=&\frac{R_{\Lambda(1520)}'~
\mathrm{or}~R_{\Lambda(1690)}'}{R_{\Lambda(1520)}'+R_{\Lambda(1690)}'}.
\end{eqnarray}

\begin{figure}[ht]
\centering \epsfxsize=9.0 cm \epsfbox{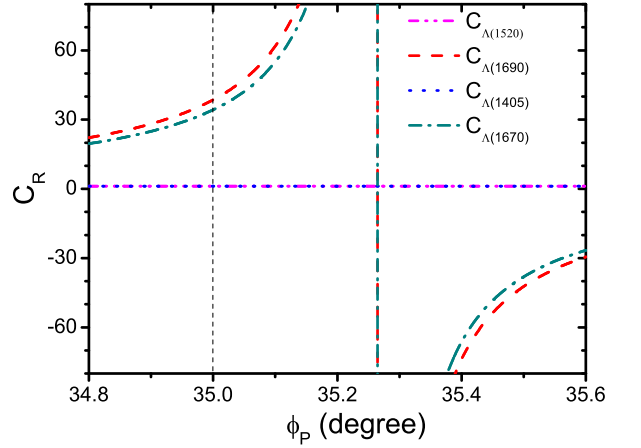} \caption{The
coupling strength parameter, $C_R$, as a function of the
$\eta$-$\eta'$ mixing angle $\phi_P$. } \label{fig-cr}
\end{figure}

With these extracted $g_R$ and $g_R'$ parameters, the  $C_R$
parameters can be easily worked out according to Eq.~(\ref{cp}). It
is found that $C_R$ is a function of the $\eta$-$\eta'$ mixing angle
$\phi_P$, which might be in the range $\phi_P\simeq (30^\circ,
47^\circ)$~\cite{Ke:2009mn,DiDonato:2011kr}. Considering the
uncertainties of $\phi_P$, we plot $C_R$ as a function of $\phi_P$
in Fig.~\ref{fig-cr}. From the figure, one can find that the $C_R$
parameters for $\Lambda(1670)S_{01}$ and $\Lambda(1690)D_{03}$ are
sensitive to the $\eta$-$\eta'$ mixing angle $\phi_P$. When takeing
a small $\eta$-$\eta'$ mixing angle $\phi_P=35^\circ$, we obtain a
large value $C_{\Lambda(1670)}\simeq 34$ for $\Lambda(1670)S_{01}$,
which can naturally explain the large contributions of
$\Lambda(1670)S_{01}$ found in the $K^-P\rightarrow\Lambda\eta$
process.

Using the determined $\eta$-$\eta'$ mixing angle $\phi_P=35^\circ$,
we also obtain a large value of $C_{\Lambda(1690)}\simeq 39$ for
$\Lambda(1690)D_{03}$. It is should be mentioned that although the
configuration mixing effects have largely enhanced the contribution
of $\Lambda(1690)$ in the $K^-P\rightarrow\Lambda\eta$ process
(about a factor of $39$), the contribution of $\Lambda(1690)D_{03}$
in the reaction is still negligibly small for the very weak coupling
to $\eta \Lambda$.

As a whole, the configuration mixing effects are crucial to
understand the strong decay properties of the low-lying negative
$\Lambda$ resonances. These resonances are most likely mixed states
between different configurations, which is consistent with the
predictions in large $N_c$ QCD~\cite{Schat:2001xr}. Considering
configuration mixing effects, we can reasonably explain the weak
coupling of $\Lambda(1670)S_{01}$ to $\bar{K}N$ and strong coupling
of $\Lambda(1670)S_{01}$ to $\eta \Lambda$, and the large strength
parameter $C_{\Lambda(1670)}\simeq 34$. The contribution of
$\Lambda(1690)D_{03}$ to the $K^-P\rightarrow\Lambda\eta$ process is
too small to give a bowl structure in the DCS, even we consider the
configuration mixing effects in these $D$-wave states.

\section{Summary}\label{summ}

In this work, we have studied the low energy reaction
$K^-p\rightarrow\Lambda\eta$ with a chiral quark model approach. A
reasonable description of the measurements has been achieved. It is
found that $\Lambda(1670)S_{01}$ dominates the reaction around at
the low energy regions, and the $t$- and $u$-channel backgrounds
also play crucial roles. Slight contributions of
$\Lambda(1405)S_{01}$ are found, however, $\Lambda(1405)S_{01}$ does
not obviously affect the shapes of the differential cross sections.
No obvious roles of the $D$-wave states $\Lambda(1520)D_{03}$ and
$\Lambda(1690)D_{03}$ are found in the reaction.

Furthermore, by the study of the $K^-p\rightarrow\Lambda\eta$
process, we have extracted the strong interaction properties of
$\Lambda(1670)S_{01}$. We find that a much large amplitude of
$\Lambda(1670)S_{01}$ in the reaction is needed, which is about 34
times (i.e., $C_{S_{01}(1670)}\simeq 34 $) larger than that derived
from the symmetry quark model. Combing this with our previous study
in~~\cite{Zhong:2013oqa}, we conclude that $\Lambda(1670)S_{01}$
should have a much weaker coupling to $\bar{K}N$, while a much
stronger coupling to $\eta \Lambda$ than that predicted in the
symmetry quark model.

To understand these strong interaction properties of
$\Lambda(1670)S_{01}$, we further study the strong decay properties
of the low-lying negative parity $\Lambda$-resonances. It is found
that the configuration mixing effects are crucial to understand the
strong decay properties of the low-lying negative $\Lambda$
resonances. These resonances are most likely mixed states between
different configurations. Considering configuration mixing effects,
we can reasonably explain the strong interaction properties of
$\Lambda(1670)S_{01}$  extracted from the
$K^-P\rightarrow\Lambda\eta$.

The data of the $K^-p\rightarrow\Lambda\eta$ process show that there
seems to exist a bowl  structure in the DCS in a narrow energy
region near the $\eta\Lambda$ threshold, which indicates a strong
$D$-wave contribution there. However, the contribution of
$\Lambda(1690)D_{03}$ to $K^-P\rightarrow\Lambda\eta$ process too
small to give a bowl structure in the DCS. Although with the
configuration mixing effects in these $D$-wave states, the amplitude
of $\Lambda(1690)D_{03}$ in the reaction could be enhanced a factor
of $\sim38$, the contribution of $\Lambda(1690)D_{03}$ is still tiny
for the very weak coupling of $\Lambda(1690)D_{03}$ to
$\eta\Lambda$. Based on the bowl structures in the DCS, Liu and Xie
believed there might exist a exotic $D$-wave state
$\Lambda(1669)D_{03}$ with a very narrow width of $\Gamma=1.5$ MeV.
To clarify whether there are contributions of a narrow $D$-wave
state or not, more accurate measurements are needed.

As a byproduct, we also have predicted the strong decay properties
of the unestablished $D$-wave state
$|\Lambda\frac{3}{2}^-\rangle_3$. This resonance mainly decay into
$\Sigma(1385)\pi$ and $\Sigma\pi$ channels. We hope the
experimentalists can search this missing $D$-wave state in the
$\Sigma(1385)\pi$ and $\Sigma\pi$ channels.

%%%%%%%%%%%%%%%%%%%%%%%%%%%%%%%%%%%%%%%%%%%%%%%%%%%%%%%%%%%%%%%%%%%%%

\section*{  Acknowledgements }
This work is supported, in part, by the National Natural Science
Foundation of China (Grants No. 11075051 and No. 11375061), Program
for Changjiang Scholars and Innovative Research Team in University
(PCSIRT, Grant No. IRT0964), the Program Excellent Talent Hunan
Normal University, the Hunan Provincial Natural Science Foundation
(Grants No. 11JJ7001 and No. 13JJ1018), and the Hunan Provincial
Innovation Foundation For Postgraduate.
%\appendix
%%%%%%%%%%%%%%%%%%%%%%%%%%%%%%%%%%%%%%%%%%%%%%%%%%%%%%%%%%%%%%%%%%555

\end{document}